\documentclass[twocolumn,floatfix,superscriptaddress,pra,aps,10pt,nofootinbib,longbibliography]{revtex4-2}
\usepackage{times}
\usepackage{amssymb}
\usepackage{color}
\usepackage{amsmath}
\usepackage{amsbsy}
\usepackage{amsthm}
\usepackage{graphicx}
\usepackage{bbm}
\usepackage{bm}
\usepackage{epsfig}
\usepackage{xfrac}
\usepackage{xcolor}
\usepackage{enumerate}
\usepackage{multirow}
\usepackage{float}
\usepackage[T1]{fontenc}

\definecolor{myurlcolor}{rgb}{0,0,0.7}
\definecolor{myrefcolor}{rgb}{0.8,0,0}
\usepackage[unicode=true,pdfusetitle, bookmarks=true,bookmarksnumbered=false,bookmarksopen=false, breaklinks=false,pdfborder={0 0 0},backref=false,colorlinks=true, linkcolor=myrefcolor,citecolor=myurlcolor,urlcolor=myurlcolor]{hyperref}

\renewcommand{\t}[1]{\textrm{#1}}

\definecolor{brightmaroon}{rgb}{0.76, 0.13, 0.28}
\definecolor{carmine}{rgb}{0.59, 0.0, 0.09}

\newcommand{\ket}[1]{|#1\rangle}

\newcommand{\bra}[1]{\langle#1|}
\newcommand{\braket}[1]{\langle #1 \rangle}

\usepackage{dsfont}

\usepackage{dsfont}

\newcommand{\bk}{{\boldsymbol{k}}}

\newcommand{\bs}[1]{{\boldsymbol{#1}}}

\begin{document}
\title{Simple and consistent spontaneous emission rate derivation\\ with a physically justified frequency cutoff}
\author{Wojciech G{\'{o}}recki}
\affiliation{Faculty of Physics, University of Warsaw, Pasteura 5, 02-093 Warsaw, Poland}
\begin{abstract}
The exact determination of the spontaneous emission coefficient for an excited atom is an extremely complex problem so various approximations are typically used. One of the most popular ones is the use of the dipole approximation of a two-level atom followed by rotating wave approximation (RWA). However, such an approach applied to the entire frequency spectrum results in the appearance of divergent integrals in the derivations, which are not treated rigorously in typical student textbooks. It is known from the literature that the introduction of cutoff for frequencies, justified by the finite size of the atom, may solve this problem. For didactic purposes, in this paper, we introduce a mathematically simple cutoff, which allows for a straightforward yet mathematically consistent rederivation of the Weisskopf-Wigner spontaneous emission rate (up to the small correction) within RWA. Importantly, this cutoff is not a mathematical trick to make calculations easier but is related to a real feature of the physical system, the neglect of which leads to inconsistency. More precise analysis demand going beyond RWA and dipole approximation.
\end{abstract}
\maketitle


\section{Introduction}
It is known that an excited atom tends to deexcite to the ground state by emitting a photon, where the probability that the atom is still in the excited state decreases (to a good approximation) exponentially with time with a $\Gamma$ factor. The formal description of this phenomenon was one of the significant challenges at the beginning of quantum mechanics \cite{weinberg1995quantum} and has been comprehensively analyzed from various perspectives over the following decades \cite{Agarwal1974}. 
In many student textbooks, this effect is derived in a significantly simplified way~\cite{scully1999quantum,sargent1974laser,meystre2007elements}, which, unfortunately, leads to some internal contradictions. The aim of this paper is to discuss and fix this inconsistency while keeping the didactic simplicity of the approximated derivation.

In deriving the effect, one may start from the dipole approximation, assuming that the size of the atom under consideration is much smaller than the electromagnetic wavelength in the range of interest, which allows us to write down the equations corresponding to spontaneous emission in a simplified version. However, the determination of the emission coefficient requires the summation of contributions from all wavelengths. Doing so without taking into account that for very short wavelengths the dipole approximation no longer works leads to divergent integrals.

This problem is not discussed in typical textbooks~\cite{scully1999quantum,sargent1974laser,meystre2007elements}, where authors write down equations derived in this way (whose strict solution would not correspond to any physically relevant situation), and then in the solution process postulate a series of unjustified approximations to recover the desired result. While the text commentary accompanying these transitions correctly captures the physical aspects associated with these steps, it does not allow us to understand the lack of consistency at the level of the equations; nor does it provide insight on what time scales we should expect consistency with the final (approximate) result.

\begin{figure}
    \centering
    \includegraphics[width=0.47\textwidth]{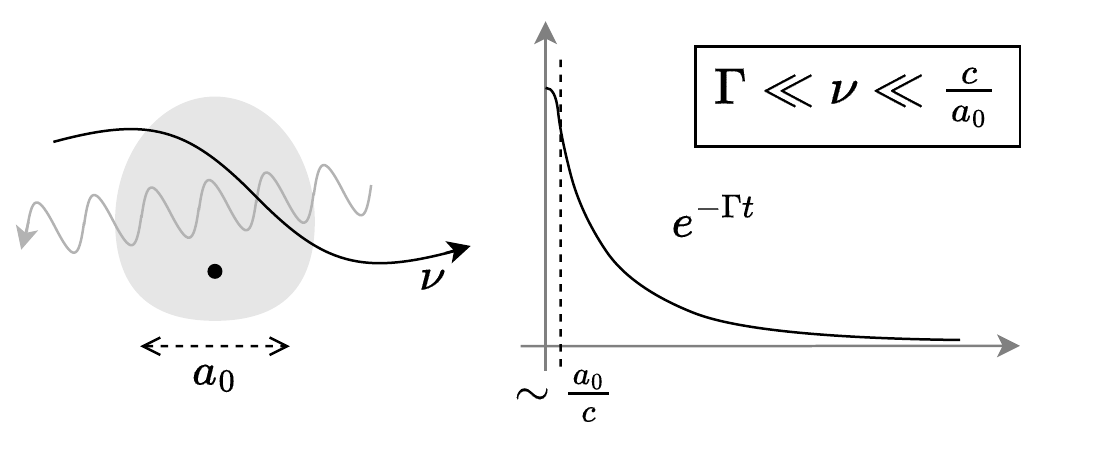}
    \caption{Dipolar atom of size $a_0$ interacts with electromagnetic waves with resonance frequency $\nu$. 
    For the light waves of a length much smaller than $a_0$, an interaction of the wave with atom averages to $0$. Starting from the excited state, the atom is likely to relax by emitting a photon, such that for times $t\gg a_0/c$ the probability of remaining in the excited state decreases exponentially with coefficient $\Gamma$.}
    \label{fig:fun}
\end{figure}

To resolve this problem, without losing the simplicity of formulas, we propose a smooth simple frequency cutoff at the physically motivated range of order $\sim c/a_0$ (where $a_0$ is Bohr radius and $c$ is the speed of light). The shape of the cutoff is chosen in a way to make the integral appearing later easy to calculate. A similar analysis has been done in \cite{seke1988deviation}, involving a noticeably more complicated mathematical formalism.

The paper is organized as follows. In section \ref{sec:standard} we recall the standard approach leading to the equations mentioned above. In section \ref{sec:cutoff} we introduce the cutoff and derive an approximate solution for times $t\gg a_0/c$. We also discuss in detail the approximations made in \cite{scully1999quantum,sargent1974laser,meystre2007elements} and conclude the crucial differences. In section \ref{sec:con} we summarize the results, comparing them with similar approaches appearing in literature, and discussing the advantages and limitations of the used method.

For completeness, in appendices we recall a derivation of electric dipole approximation, starting from full Hamiltonian of a non-relativistic charged particle interacting with quantized electromagnetic field (App. \ref{app:derivation}), discuss broader the concept of smooth cutoff (App. \ref{app:smooth}) and stress the advantage of using $\bs E\cdot \bs r$ interaction Hamiltonian over $\bs A\cdot \bs p$ for the discussed problem (App. \ref{app:aper}).

\section{Standard derivation of spontaneous emission rate}
\label{sec:standard}

The evolution of a single electrically neutral atom interacting with an electromagnetic field is typically approximated by the Hamiltonian:
\begin{equation}
    H=H_A+ H_L+ H_I,
\end{equation}
where $ H_A=\sum_i \hbar\nu_i \ket{i}\bra{i}$ is the Hamiltonian of the atom, $ H_L=\sum_{\bs k,s} \hbar \omega_{\bs k} \hat a_{\bs k,s}^\dagger \hat a_{\bs k,s}$ is the Hamiltonian of the light (where $\hat a_{\bs k,s}$ is an annihilator operator of the plane wave of wave vector $\bs k$ and polarization $s$) and $H_I$ is the interaction Hamiltonian. For conceptual simplicity, we model infinite free-dimensional space by a cube box of size $V$, so $\bs k$ takes only discrete values.

Assuming, that we are interested mainly in the light waves of a length much bigger than the size of atom $\frac{1}{|\bs k|}\gg a_0$, it is reasonable to approximate the value of the electromagnetic field by its value at point zero $\bs E(\bs r)\approx \bs E(0)$, to get:
\begin{equation}
    H_I=-\bs d\cdot\bs E(0),\quad \bs E(0)=\sum_{\bk,s}\sqrt{\frac{\hbar\omega_\bk}{2\epsilon_0 V}}(\hat a_{\bk,s}+\hat a_{\bk,s}^\dagger)\bs e_{\bk,s},
\end{equation}
where $\bs d=-e \bs r$ is electric dipole moment, while $\bs e_{\bk,s}$ is a unit polarization vector (perpendicular to $\bs k$). We restrict to only two levels of the atom -- the ground $\ket{g}$ and excited one $\ket{e}$, for which $\bs d_{eg}:=\braket{e|\bs d|g}$ (while $\braket{e|\bs d|e}=\braket{g|\bs d|g}=0$) and $\nu=\nu_e-\nu_g$. By applying rotating wave approximation (RWA), in the Dirac picture, we obtain:
\begin{equation}
\label{eq:dirac}
    H_I^D(t)=\hbar\sum_{\bk,s}g_{\bk,s}\ket{e}\bra{g}\hat a_{\bk,s}e^{i(\nu-\omega_{\bk})t}+h.c.
\end{equation}
with $g_{\bs k,s}=-\frac{\bs e_{\bs k,s}\cdot \bs d_{eg}}{\hbar}\sqrt{\frac{\hbar\omega_\bk}{2\epsilon_0V}}$. As in RWA the total number of excitation is conserved, for all $t$ the state has a form:
\begin{equation}
\ket{\psi(t)}=c_e(t)\ket{e}\otimes\ket{0}+\sum_{\bk,s} c_{g;\bs k,s}(t)\ket{g}\otimes \ket{1}_{\bs k,s}.
\end{equation}
Using $\ket{\dot\psi(t)}=-\frac{i}{\hbar}H_I^D(t)\ket{\psi(t)}$ we obtain:
\begin{equation}
\begin{cases}
    \dot c_e(t)&=-i \sum g_{\bs k,s} e^{i(\nu-\omega_k)t}c_{g;\bs k,s}(t),\\
    \dot c_{g;\bs k,s}(t)&=-i g^*_{\bs k,s} e^{-i(\nu-\omega_k)t}c_e(t).
    \end{cases}
\end{equation}
To solve it we integrate the latter one over $\int_0^t d t'$ and substitute it to the first one. Assuming $V$ is large, we approximate the sum by the proper integral:
\begin{equation}
\label{eq:cont}
\sum_{\bk,s}\to 2\frac{V}{(2\pi)^3}\int_0^{2\pi}d\varphi \int_0^\pi d\theta \sin(\theta) \int_0^\infty dk k^2,
\end{equation}
where factor $2$ at the beginning comes from summation over polarizations $s$. Following notation from \cite{scully1999quantum}, the angles $(\theta,\varphi)$ in the integral corresponds to the orientation of the vectors $\bs e_{\bk,s}$ (not $\bk$), so $(\bs e_{\bk,s} \cdot \bs d_{eg})^2=|\bs d_{eg}|^2\cos^2\theta $, and integral over $\theta$ gives $\int_0^\pi d\theta \sin\theta \cos^2\theta=\frac{2}{3}$. Finally we obtain integro-differential equation for $c_e(t)$:
\begin{equation}
\label{eq:main0}
    \dot c_e(t)=-D \int_0^\infty d\omega \omega^3 \int_0^t e^{i(\nu-\omega)(t-t')}c_e(t')dt',
\end{equation}
with $D=\frac{2|d_{eg}|^2}{3(2\pi)^2\epsilon_0 \hbar c^3}$, which is the one appearing in \cite{scully1999quantum,sargent1974laser,meystre2007elements}. However, as stated in the introduction, solving this equation does not allow for the recovery of the well-known results involving an exponential decay of the probability of the atom remaining in the excited state, which we will discuss in detail at the end of the next section.

\section{Frequency cutoff}
\label{sec:cutoff}

To take into account the fact, that for the waves of frequencies $\omega\gg c/a_0$ the electromagnetic field averages to zero on the size of the atom, we introduce the cutoff correction $e^{-\epsilon \omega}$ to \eqref{eq:main0} (see App. \ref{app:derivation} for more detail discussion of the derivation of dipole approximation):
\begin{equation}
\label{eq:main}
    \dot c_e(t)=-D \int_0^\infty d\omega \omega^3 
e^{-\epsilon \omega}\int_0^t e^{i(\nu-\omega)(t-t')}c_e(t')dt',
\end{equation}
where $\epsilon$ is supposed to be of order $a_0/c$ (which, for the actual atoms, implies $\epsilon\nu\ll 1$). Let us introduce $\tau=t-t'$. The first integral is solvable analytically (by applying integration by parts three times):
\begin{equation}
\label{eq:fun}
    \int_0^\infty d\omega \omega^3 e^{-\epsilon \omega}e^{-i\omega\tau}=\frac{6}{(\tau-i \epsilon)^4}.
\end{equation}
The function \eqref{eq:fun} is concentrated mainly in a close neighborhood of $\tau=0$ (see Fig. \ref{fig:fun}). Therefore, assuming that $c_e(t-\tau)$ changes much slower than $e^{i\nu\tau}$, namely:
\begin{equation*}
(*)\quad  \frac{|\dot c_e|}{|c_e|}\ll \nu,
\end{equation*}
we may approximate:
\begin{equation}
\label{eq:intfun}
    \int_0^t d\tau \frac{6e^{i\nu\tau}c_e(t-\tau)}{(\tau-i \epsilon)^4}\overset{(*)}{\approx} c_e(t)\cdot  \int_0^t d\tau \frac{6e^{i\nu\tau}}{(\tau-i \epsilon)^4}.
\end{equation}
The assumption $(*)$ is crucial for further reasoning. Therefore, after obtaining the final solution we need to go back here to check if it truly satisfies it. 

\begin{figure}
    \centering
    \includegraphics[width=0.4\textwidth]{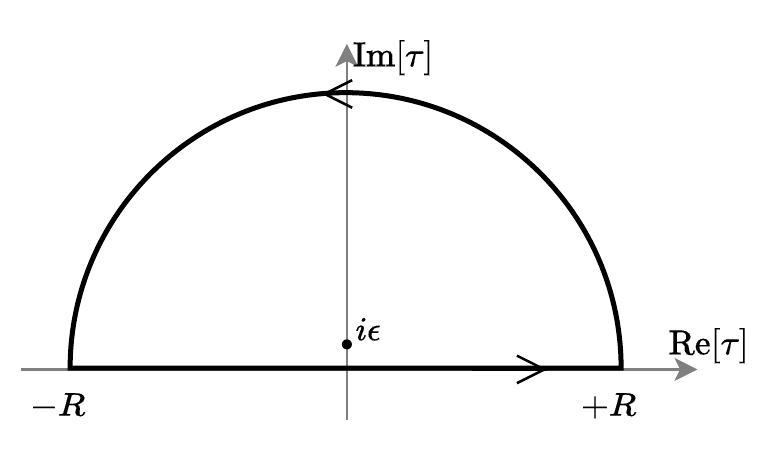}
    \caption{The contour used in calculating \eqref{eq:residuum}.}
    \label{fig:residuum}
\end{figure}

With above approximation, Eq. \eqref{eq:main} takes the form $\frac{d}{dt}c_e=c_e\cdot (a+ib)$ (with $a,b\in\mathbb R$). Therefore the decay of $|c_e(t)|^2$ depends only on the real element $\frac{d}{dt}|c_e|^2=2a|c_e|^2$, so we will be interested mainly in the real part of integral \eqref{eq:intfun}.
\begin{figure}
    \centering
    \includegraphics[width=0.4\textwidth]{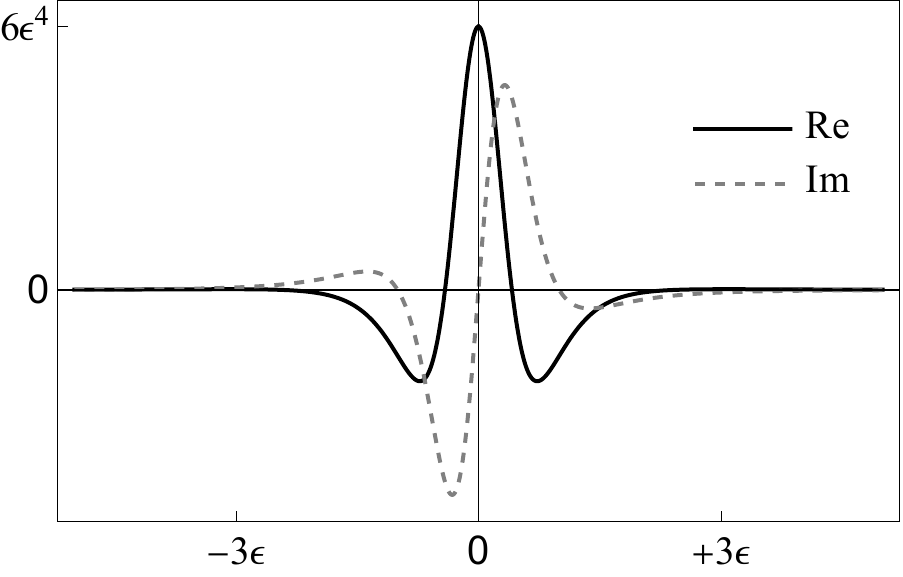}
    \caption{Real and imagine parts of function \eqref{eq:fun}. Both are strongly concentrated in $\epsilon$-size neighborhood of $0$.}
    \label{fig:fun}
\end{figure}
As $\epsilon\nu\ll1$, for $t\gg \epsilon$ the integral \eqref{eq:intfun} will not change significantly after extending range $t\to\infty$. Next, note that both $e^{i \nu \tau}$ and $6/(\tau-i\epsilon)^4$ have symmetric real parts and anti-symmetric imaginary parts (see Fig. \ref{fig:fun}). Therefore also their product has a symmetric real part and an anti-symmetric imaginary part, so:
\begin{equation}
     \t{Re}\left[\int_0^\infty d\tau \frac{6e^{i\nu\tau}}{(\tau-i \epsilon)^4}\right]= \frac{1}{2}\int_{-\infty}^{+\infty} d\tau \frac{6e^{i\nu\tau}}{(\tau-i \epsilon)^4}.
\end{equation}
To calculate it we use Cauchy integral formula:
\begin{equation}
    \oint \frac{f(z)}{(z-z_0)^{n+1}}=\frac{2\pi i}{n!}f^{(n)}(z_0)
\end{equation}
with $z_0=i\epsilon$. We choose the contour which goes from $-R$ to $R$ on the real line and then along half or circle in the upper part of the complex plane ($\t{Im}\tau\geq 0$, see Fig. \ref{fig:residuum}). Taking $R\to\infty$, the integral over half of the circle goes to $0$, so we have:
\begin{equation}
\label{eq:residuum}
    \frac{1}{2}\int_{-\infty}^{+\infty} d\tau \frac{6e^{i\nu\tau}}{(\tau-i \epsilon)^4}=\frac{1}{2}\frac{2\pi i}{6}6(i\nu)^3e^{-\nu\epsilon}=\pi \nu^3e^{-\nu\epsilon}.
\end{equation}
Introducing $\Gamma=2\pi\nu^3 D$ and using formula $\frac{d}{dt}|c_e|^2=2a|c_e|^2$, we end with:
\begin{equation}
    \frac{d}{dt} |c_e(t)|^2\overset{(*)}{\approx} -\Gamma e^{-\nu\epsilon}|c_e(t)|^2.
\end{equation}
At this point, one could naively think that with $\epsilon\to 0$, the effect of cutoff became irrelevant and we recover standard decay rate $\Gamma$. However, we should still remember to check the condition $(*)$ used in \eqref{eq:intfun}. It requires that not only $\Gamma\ll\nu$, but also an imaginary part of integral from \eqref{eq:intfun} need to be much smaller than $\nu$.

During calculating the imaginary part we may be much less subtle than previously. We may approximate the whole integral with reasonable precision by
\begin{multline}
    \int_0^\infty d\tau \frac{6e^{i\nu\tau}}{(\tau-i \epsilon)^4}\overset{u=\tau/\epsilon}{=} \frac{1}{\epsilon^3}\int_0^\infty du \frac{6e^{i\epsilon\nu u}}{(u-i)^4}\\
    \overset{\epsilon\nu\ll 1}{\approx} \frac{1}{\epsilon^3}\int_0^\infty du \frac{6}{(u-i)^4}=-\frac{2i}{\epsilon^3},
\end{multline}
where in the second step we used the fact, that the main body of the integral is accumulated in the distance of order $\sim 1$ from the origin, where the function $e^{i\epsilon\nu u}$ is approximately constant (as $\epsilon\nu\ll 1$).

Note, that such an approximation leads to a purely imaginary result 
(which causes a phase correction, related to Lamb shift), completely neglecting the real part. It is because the real part is smaller of an order of magnitude $\sim (\epsilon\nu)^3$ and its existence result from a non-zero derivative of $e^{i\epsilon\nu u}$. However, unlike the imaginary part, the impact of the real part accumulates during evolution, so for a long time it plays a crucial role, so it cannot be omitted.

We therefore end with:
\begin{equation}
\label{eq:exact}
    \dot c_e(t)\overset{(*)}{\approx} \left[-\Gamma/2 \cdot e^{-\nu\epsilon}+i 2D/\epsilon^3)\right]c_e(t),
\end{equation}
where to satisfy $(*)$ it is needed $2D/\epsilon^3\ll \nu$. To recover the standard result we need also $e^{-\epsilon\nu}\approx 1$. For hydrogen atom eigenstates $1s$ and $2p$ electric dipole moment is equal $|\bs d_{eg}|=\frac{128\sqrt{2}}{243}ea_0\approx \frac{3}{4}ea_0$, which gives $D\approx\frac{\alpha}{2\pi}\left(\frac{a_0}{c}\right)^2$ (where $\alpha$ is fine structure constant). As $\nu=\frac{3\alpha}{8}\frac{c}{a_0}$, above condition simplifies to:
\begin{equation}
\sqrt[3]{\frac{8}{3\pi}}\ll \epsilon \frac{c}{a_0} \ll \frac{8}{3\alpha},
\end{equation}
which may be satisfied by taking $\epsilon\approx 10\frac{a_0}{c}$. This allows us to recover the well-known result -- for times $t\gg \epsilon$:
\begin{equation}
\label{eq:result}
    |c_e(t)|^2\approx e^{-\Gamma t},
\end{equation}
what was to be shown.

The magnitude of corrections appearing in \eqref{eq:exact} (compering to \eqref{eq:result}) should be treated with severely limited trust, as the model, we used in \eqref{eq:dirac}, neglects a lot of effects of similar importance. Still, we have shown that even in such a simplified approach, the frequency cutoff is necessary for obtaining reasonable results.

Let us now briefly comment on derivations appearing in some of the standard textbooks. In \cite{scully1999quantum}, in \eqref{eq:main} at the beginning the $\omega^3$ is replaced by $\nu^3$ under the integral, and then the integral is expanded to $\omega\in (-\infty,+\infty)$ with comments, that such changes do not affect the result. While $\Gamma$ obtained this way is the same, such an approach results in a lack of correction to frequency. That shows clearly, that while the equations proposed by \cite{scully1999quantum} correctly describe the physical process of amplitude decaying, mathematically it is not an approximation of the initial equation \eqref{eq:main}. 

Alternatively, in \cite{sargent1974laser,meystre2007elements}, by applying some heuristic arguments, the authors obtain the results qualitatively similar to the one obtained here for $\epsilon \to 0$. This, however, leading to infinite Lamb shift, simply states in contradiction with an assumption about slowly varying of $c(t)$ (which was also assumed in \cite{sargent1974laser,meystre2007elements}), which makes derivation not self-consistent.

\section{Broader discussion and conclusions}
\label{sec:con}

As stated before, the shape of the cutoff function has been chosen here arbitrarily, to make calculations easier. More detailed analysis of interacting electromagnetic field with hydrogen atom has been performed in \cite{moses1973photon}, where Coulombs gauge has been used $\bs \nabla \cdot \bs A(\bs r, t)=0$, so the interaction Hamiltonian takes the form $\frac{e}{m}\bs A\cdot \bs p$. The direct calculation (where, instead of plane-waves basis of photons, the basis with well-defined energy, total angular momentum, magnetic quantum number, and helicity has been used) shows that for such an approach the term $\omega^3$ in \eqref{eq:main0} should be replaced by:
\begin{equation}
\label{eq:cutoff}
\frac{\nu^2\omega}{(1+\epsilon^2\omega^2)^4},\quad \t{with}\quad \epsilon=\frac{2}{3}\frac{a_0}{c}.
\end{equation}
The calculation of spontaneous emission rate with the cutoff \eqref{eq:cutoff}, as well as with a sharp cutoff, has been performed in \cite{seke1988deviation}, with the usage of Laplace transformation. In precise analysis, the authors derived the correction to exponential decay for the larger times $t$. Namely, it was shown that for large $t$ the amplitude of the excited state decreases slower than exponentially. Broader analysis suggests that this feature is independent of the exact shape of the cutoff~\cite{BERMAN2010175}. We can see a clue to the existence of this effect in the above derivation -- looking at approximation \eqref{eq:intfun} one easily sees that it brakes down when the ratio $c_e(t)/c_e(0)$ becomes extremely small (as at some point $c_e(t)/\epsilon^4$ became smaller than $c_e(0)/t^4$).

Note, that while close to resonance formula \eqref{eq:cutoff} is approximately equal to $\approx \nu^3$, for the small frequency, it goes to $\nu^2\omega$ instead of $\omega^3$. That is a straight consequence of using interaction Hamiltonian $\frac{e}{m}\bs A\cdot \bs p$ instead of $e\bs E\cdot \bs r$ (which are equivalent if full calculations including whole Hilbert space is performed but lead to different results within two-level approximation; see App. \ref{app:aper} or \cite{cohen1997photons} for broad discussion). 
In general, the form $e\bs E\cdot \bs r$ is safer for analyzing within two-level approximation, especially in the low-frequency regime \cite{rzazewski2004equivalence}. Therefore, the exact values of the corrections obtained using \eqref{eq:cutoff} should also be treated with limited trust.

The clue message of this paper is that to connect the \eqref{eq:main} with well-known results \eqref{eq:result}, the cutoff needs to be simultaneously sufficiently weak to not affect significantly the rate of decay $\Gamma$ and strong enough to make the frequency correction small. The fact, that this condition meets the results coming from the actual size of the atom justified the whole procedure.
More accurate quantitative analysis of corrections would require going beyond RWA and two-level atom approximation to full relativistic quantum electrodynamics theorem, including i.a. self electron energy and renormalization \cite{weinberg1995quantum2}; see \cite{allen1981rezonans,zheng2008quantum,li2012collective} for examples of a broader analysis.

\textit{Acknowledgements.} The Author thanks Rafa{\l} Demkowicz-Dobrza{\'{n}}ski, Krzysztof Paw{\l}owski and Stanis{\l}aw Kurdziałek for comments on the clarity of the paper and also Kazimierz Rz\k{a}\.zewski for the discussion on the subject.

\bibliography{biblio}

\begin{thebibliography}{14}%
\makeatletter
\providecommand \@ifxundefined [1]{%
 \@ifx{#1\undefined}
}%
\providecommand \@ifnum [1]{%
 \ifnum #1\expandafter \@firstoftwo
 \else \expandafter \@secondoftwo
 \fi
}%
\providecommand \@ifx [1]{%
 \ifx #1\expandafter \@firstoftwo
 \else \expandafter \@secondoftwo
 \fi
}%
\providecommand \natexlab [1]{#1}%
\providecommand \enquote  [1]{``#1''}%
\providecommand \bibnamefont  [1]{#1}%
\providecommand \bibfnamefont [1]{#1}%
\providecommand \citenamefont [1]{#1}%
\providecommand \href@noop [0]{\@secondoftwo}%
\providecommand \href [0]{\begingroup \@sanitize@url \@href}%
\providecommand \@href[1]{\@@startlink{#1}\@@href}%
\providecommand \@@href[1]{\endgroup#1\@@endlink}%
\providecommand \@sanitize@url [0]{\catcode `\\12\catcode `\$12\catcode
  `\&12\catcode `\#12\catcode `\^12\catcode `\_12\catcode `\%12\relax}%
\providecommand \@@startlink[1]{}%
\providecommand \@@endlink[0]{}%
\providecommand \url  [0]{\begingroup\@sanitize@url \@url }%
\providecommand \@url [1]{\endgroup\@href {#1}{\urlprefix }}%
\providecommand \urlprefix  [0]{URL }%
\providecommand \Eprint [0]{\href }%
\providecommand \doibase [0]{https://doi.org/}%
\providecommand \selectlanguage [0]{\@gobble}%
\providecommand \bibinfo  [0]{\@secondoftwo}%
\providecommand \bibfield  [0]{\@secondoftwo}%
\providecommand \translation [1]{[#1]}%
\providecommand \BibitemOpen [0]{}%
\providecommand \bibitemStop [0]{}%
\providecommand \bibitemNoStop [0]{.\EOS\space}%
\providecommand \EOS [0]{\spacefactor3000\relax}%
\providecommand \BibitemShut  [1]{\csname bibitem#1\endcsname}%
\let\auto@bib@innerbib\@empty
\bibitem [{\citenamefont {Weinberg}(1995{\natexlab{a}})}]{weinberg1995quantum}%
  \BibitemOpen
  \bibfield  {author} {\bibinfo {author} {\bibfnamefont {S.}~\bibnamefont
  {Weinberg}},\ }\href@noop {} {\emph {\bibinfo {title} {The quantum theory of
  fields}}},\ Vol.~\bibinfo {volume} {1}\ (\bibinfo  {publisher} {Cambridge
  university press},\ \bibinfo {year} {1995})\ Chap.\ \bibinfo {chapter}
  {1.2}\BibitemShut {NoStop}%
\bibitem [{\citenamefont {Agarwal}(1974)}]{Agarwal1974}%
  \BibitemOpen
  \bibfield  {author} {\bibinfo {author} {\bibfnamefont {G.~S.}\ \bibnamefont
  {Agarwal}},\ }\bibinfo {title} {Quantum statistical theories of spontaneous
  emission and their relation to other approaches},\ in\ \href
  {https://doi.org/10.1007/BFb0042382} {\emph {\bibinfo {booktitle} {Quantum
  Optics}}},\ \bibinfo {editor} {edited by\ \bibinfo {editor} {\bibfnamefont
  {G.}~\bibnamefont {H{\"o}hler}}}\ (\bibinfo  {publisher} {Springer Berlin
  Heidelberg},\ \bibinfo {address} {Berlin, Heidelberg},\ \bibinfo {year}
  {1974})\ pp.\ \bibinfo {pages} {1--128}\BibitemShut {NoStop}%
\bibitem [{\citenamefont {Scully}\ and\ \citenamefont
  {Zubairy}(1999)}]{scully1999quantum}%
  \BibitemOpen
  \bibfield  {author} {\bibinfo {author} {\bibfnamefont {M.~O.}\ \bibnamefont
  {Scully}}\ and\ \bibinfo {author} {\bibfnamefont {M.~S.}\ \bibnamefont
  {Zubairy}},\ }\href@noop {} {\bibinfo {title} {Quantum optics}} (\bibinfo
  {year} {1999})\BibitemShut {NoStop}%
\bibitem [{\citenamefont {Sargent~III}\ \emph {et~al.}(1974)\citenamefont
  {Sargent~III}, \citenamefont {Scully},\ and\ \citenamefont
  {Lamb~Jr}}]{sargent1974laser}%
  \BibitemOpen
  \bibfield  {author} {\bibinfo {author} {\bibfnamefont {M.}~\bibnamefont
  {Sargent~III}}, \bibinfo {author} {\bibfnamefont {M.}~\bibnamefont
  {Scully}},\ and\ \bibinfo {author} {\bibfnamefont {W.}~\bibnamefont
  {Lamb~Jr}},\ }\href@noop {} {\emph {\bibinfo {title} {Laser Physic}}}\
  (\bibinfo  {publisher} {Addision-Wesley Publishing Company},\ \bibinfo {year}
  {1974})\BibitemShut {NoStop}%
\bibitem [{\citenamefont {Meystre}\ and\ \citenamefont
  {Sargent}(2007)}]{meystre2007elements}%
  \BibitemOpen
  \bibfield  {author} {\bibinfo {author} {\bibfnamefont {P.}~\bibnamefont
  {Meystre}}\ and\ \bibinfo {author} {\bibfnamefont {M.}~\bibnamefont
  {Sargent}},\ }\href@noop {} {\emph {\bibinfo {title} {Elements of quantum
  optics}}}\ (\bibinfo  {publisher} {Springer Science \& Business Media},\
  \bibinfo {year} {2007})\BibitemShut {NoStop}%
\bibitem [{\citenamefont {Seke}\ and\ \citenamefont
  {Herfort}(1988)}]{seke1988deviation}%
  \BibitemOpen
  \bibfield  {author} {\bibinfo {author} {\bibfnamefont {J.}~\bibnamefont
  {Seke}}\ and\ \bibinfo {author} {\bibfnamefont {W.~N.}\ \bibnamefont
  {Herfort}},\ }\bibfield  {title} {\bibinfo {title} {Deviations from
  exponential decay in the case of spontaneous emission from a two-level
  atom},\ }\href {https://doi.org/10.1103/PhysRevA.38.833} {\bibfield
  {journal} {\bibinfo  {journal} {Phys. Rev. A}\ }\textbf {\bibinfo {volume}
  {38}},\ \bibinfo {pages} {833} (\bibinfo {year} {1988})}\BibitemShut
  {NoStop}%
\bibitem [{\citenamefont {Moses}(1973)}]{moses1973photon}%
  \BibitemOpen
  \bibfield  {author} {\bibinfo {author} {\bibfnamefont {H.~E.}\ \bibnamefont
  {Moses}},\ }\bibfield  {title} {\bibinfo {title} {Photon wave functions and
  the exact electromagnetic matrix elements for hydrogenic atoms},\ }\href
  {https://doi.org/10.1103/PhysRevA.8.1710} {\bibfield  {journal} {\bibinfo
  {journal} {Phys. Rev. A}\ }\textbf {\bibinfo {volume} {8}},\ \bibinfo {pages}
  {1710} (\bibinfo {year} {1973})}\BibitemShut {NoStop}%
\bibitem [{\citenamefont {Berman}\ and\ \citenamefont
  {Ford}(2010)}]{BERMAN2010175}%
  \BibitemOpen
  \bibfield  {author} {\bibinfo {author} {\bibfnamefont {P.~R.}\ \bibnamefont
  {Berman}}\ and\ \bibinfo {author} {\bibfnamefont {G.~W.}\ \bibnamefont
  {Ford}},\ }\bibfield  {title} {\bibinfo {title} {Chapter 5 - spontaneous
  decay, unitarity, and the weisskopf–wigner approximation},\ }in\ \href
  {https://doi.org/https://doi.org/10.1016/S1049-250X(10)59005-0} {\emph
  {\bibinfo {booktitle} {Advances in Atomic, Molecular, and Optical
  Physics}}},\ \bibinfo {series} {Advances In Atomic, Molecular, and Optical
  Physics}, Vol.~\bibinfo {volume} {59},\ \bibinfo {editor} {edited by\
  \bibinfo {editor} {\bibfnamefont {E.}~\bibnamefont {Arimondo}}, \bibinfo
  {editor} {\bibfnamefont {P.}~\bibnamefont {Berman}},\ and\ \bibinfo {editor}
  {\bibfnamefont {C.}~\bibnamefont {Lin}}}\ (\bibinfo  {publisher} {Academic
  Press},\ \bibinfo {year} {2010})\ pp.\ \bibinfo {pages}
  {175--221}\BibitemShut {NoStop}%
\bibitem [{\citenamefont {Cohen-Tannoudji}\ \emph {et~al.}(1997)\citenamefont
  {Cohen-Tannoudji}, \citenamefont {Dupont-Roc},\ and\ \citenamefont
  {Grynberg}}]{cohen1997photons}%
  \BibitemOpen
  \bibfield  {author} {\bibinfo {author} {\bibfnamefont {C.}~\bibnamefont
  {Cohen-Tannoudji}}, \bibinfo {author} {\bibfnamefont {J.}~\bibnamefont
  {Dupont-Roc}},\ and\ \bibinfo {author} {\bibfnamefont {G.}~\bibnamefont
  {Grynberg}},\ }\href@noop {} {\emph {\bibinfo {title} {Photons and atoms:
  introduction to quantum electrodynamics}}}\ (\bibinfo {year}
  {1997})\BibitemShut {NoStop}%
\bibitem [{\citenamefont {Rzazewski}\ and\ \citenamefont
  {Boyd}(2004)}]{rzazewski2004equivalence}%
  \BibitemOpen
  \bibfield  {author} {\bibinfo {author} {\bibfnamefont {K.}~\bibnamefont
  {Rzazewski}}\ and\ \bibinfo {author} {\bibfnamefont {R.~W.}\ \bibnamefont
  {Boyd}},\ }\bibfield  {title} {\bibinfo {title} {Equivalence of interaction
  hamiltonians in the electric dipole approximation},\ }\href
  {https://doi.org/10.1080/09500340408230412} {\bibfield  {journal} {\bibinfo
  {journal} {Journal of Modern Optics}\ }\textbf {\bibinfo {volume} {51}},\
  \bibinfo {pages} {1137} (\bibinfo {year} {2004})}\BibitemShut {NoStop}%
\bibitem [{\citenamefont
  {Weinberg}(1995{\natexlab{b}})}]{weinberg1995quantum2}%
  \BibitemOpen
  \bibfield  {author} {\bibinfo {author} {\bibfnamefont {S.}~\bibnamefont
  {Weinberg}},\ }\href@noop {} {\emph {\bibinfo {title} {The quantum theory of
  fields}}},\ Vol.~\bibinfo {volume} {2}\ (\bibinfo  {publisher} {Cambridge
  university press},\ \bibinfo {year} {1995})\BibitemShut {NoStop}%
\bibitem [{\citenamefont {Allen}\ \emph {et~al.}(1981)\citenamefont {Allen},
  \citenamefont {Eberly},\ and\ \citenamefont
  {Rz{\k{a}}{\.z}ewski}}]{allen1981rezonans}%
  \BibitemOpen
  \bibfield  {author} {\bibinfo {author} {\bibfnamefont {L.}~\bibnamefont
  {Allen}}, \bibinfo {author} {\bibfnamefont {J.}~\bibnamefont {Eberly}},\ and\
  \bibinfo {author} {\bibfnamefont {K.}~\bibnamefont {Rz{\k{a}}{\.z}ewski}},\
  }\href@noop {} {\emph {\bibinfo {title} {Rezonans optyczny}}}\ (\bibinfo
  {publisher} {Pa{\'n}stwowe Wydaw. Naukowe},\ \bibinfo {year} {1981})\ Chap.\
  \bibinfo {chapter} {11; polish extended edition of \textit{Optical resonance
  and two-level atoms}}\BibitemShut {NoStop}%
\bibitem [{\citenamefont {Zheng}\ \emph {et~al.}(2008)\citenamefont {Zheng},
  \citenamefont {Zhu},\ and\ \citenamefont {Zubairy}}]{zheng2008quantum}%
  \BibitemOpen
  \bibfield  {author} {\bibinfo {author} {\bibfnamefont {H.}~\bibnamefont
  {Zheng}}, \bibinfo {author} {\bibfnamefont {S.~Y.}\ \bibnamefont {Zhu}},\
  and\ \bibinfo {author} {\bibfnamefont {M.~S.}\ \bibnamefont {Zubairy}},\
  }\bibfield  {title} {\bibinfo {title} {Quantum zeno and anti-zeno effects:
  Without the rotating-wave approximation},\ }\href
  {https://doi.org/10.1103/PhysRevLett.101.200404} {\bibfield  {journal}
  {\bibinfo  {journal} {Phys. Rev. Lett.}\ }\textbf {\bibinfo {volume} {101}},\
  \bibinfo {pages} {200404} (\bibinfo {year} {2008})}\BibitemShut {NoStop}%
\bibitem [{\citenamefont {Li}\ \emph {et~al.}(2012)\citenamefont {Li},
  \citenamefont {Evers}, \citenamefont {Zheng},\ and\ \citenamefont
  {Zhu}}]{li2012collective}%
  \BibitemOpen
  \bibfield  {author} {\bibinfo {author} {\bibfnamefont {Y.}~\bibnamefont
  {Li}}, \bibinfo {author} {\bibfnamefont {J.}~\bibnamefont {Evers}}, \bibinfo
  {author} {\bibfnamefont {H.}~\bibnamefont {Zheng}},\ and\ \bibinfo {author}
  {\bibfnamefont {S.-Y.}\ \bibnamefont {Zhu}},\ }\bibfield  {title} {\bibinfo
  {title} {Collective spontaneous emission beyond the rotating-wave
  approximation},\ }\href {https://doi.org/10.1103/PhysRevA.85.053830}
  {\bibfield  {journal} {\bibinfo  {journal} {Phys. Rev. A}\ }\textbf {\bibinfo
  {volume} {85}},\ \bibinfo {pages} {053830} (\bibinfo {year}
  {2012})}\BibitemShut {NoStop}%
\end{thebibliography}%

\appendix

\section{Derivation of dipole approximation}
\label{app:derivation}
The Hamiltonian of a charged particle in the Coulomb potential, interacting with the electromagnetic field, is given as:
\begin{equation}
\label{eq:fullham}
H=\frac{1}{2m}(\bs p- q \bs A(\bs r))^2+V_{\t{Coul}}(\bs r)+\sum_{\bk,s} \hbar \omega_{\bk} \hat a_{\bk,s}^\dagger \hat a_{\bk,s},
\end{equation}
where the vector potential is:
\begin{equation}
\bs{A}(\bs r)=\sum_{\bk,s}\sqrt{\frac{\hbar }{2\omega_{\bk}V\epsilon_0}}\bs{e}_{\bk,s}\left(-i\hat a_{\bk,s}e^{i\bs k \bs r}+h.c.\right),
\end{equation}
so the interaction part takes the form $-\frac{q}{m}\bs A(\bs r)\cdot \bs p$. Note, that Coulomb potential is treated semi-classically here.

If the wavelength is much bigger than the size of atoms $\frac{1}{k}\gg a_0$, the field may be well approximated by its value in the origin $\bs A(0)$. To obtain a standard dipole approximation, we apply a unitary transformation
\begin{equation}
T=\exp\left[-\frac{i}{\hbar}q\bs r \cdot\bs A(0)\right],
\end{equation}
which shifts the momenta:
\begin{equation}
T \bs pT^\dagger=\bs p+\left[-\frac{i}{\hbar}q\bs r \cdot\bs A(0),\bs p\right]
=\bs p+q \bs A(0)
\end{equation}
and electromagnetic field operators:
\begin{equation}
T\hat a_{\bk,s}T^\dagger=\hat a_{\bk,s}+\lambda_{\bk,s},\quad \t{where}\quad \lambda_{\bk,s}=\frac{-q}{\sqrt{2\epsilon_0\hbar\omega_\bk V}}\bs e_{\bk,s}\cdot \bs r.
\end{equation}
It changes the last term of \eqref{eq:fullham}, $\sum_{\bk,s} \hbar \omega_{\bk,s} \hat a_{\bk,s}^\dagger \hat a_{\bk,s}$, into $
\sum_{\bk,s} \hbar \omega_{\bk,s} \hat a_{\bk,s}^\dagger \hat a_{\bk,s}+
\sum_{\bk,s} \hbar\omega_{\bk,s} \lambda_{\bk,s} (\hat a_{\bk,s}+\hat a_{\bk,s}^\dagger)
+\sum_{\bk,s} \hbar \omega_{\bk,s} \lambda_{\bk,s}^2$. Noting, that:
\begin{equation}
\bs{E}(\bs r)=\sum_{\bk,s}\sqrt{\frac{\omega_{\bk}\hbar }{2V\epsilon_0}}\bs{e}_{\bk,s}\left(\hat a_{\bk,s}e^{i\bs k \bs r}+h.c.\right),
\end{equation}
we finally obtain:
\begin{multline}
THT^\dagger=\frac{1}{2m}(\bs p-q(\bs A(\bs r)-\bs A(0))^2+V_{\t{Coul}}(\bs r)
\\-q \bs r\cdot \bs E(0)
+\frac{q^2}{2\epsilon_0}\sum_{\bs k,s} \frac{1}{V}(\bs e_{\bs k,s}\cdot \bs r)^2.
\end{multline}
We see, that the last term, which we call $E_{\t{dip}}$, does not converge. Moreover, for large $\omega$, the approximation $\bs A(\bs r)\approx\bs A(0)$ is not valid. Therefore to make use of such kind of transformation, one needs to introduce frequency cutoff. It may be done in a sharp way, by introducing some large frequency $\Omega$, defining:
\begin{equation}
\bs A^{\Omega}=\sum_{|\bk|^2\leq \Omega^2/c^2}\sqrt{\frac{\hbar }{2\omega_{\bk}V\epsilon_0}}\bs{e}_{\bk,s}\left(-i \hat a_{\bk,s}+h.c.\right)
\end{equation}
and correspoding $T^\Omega=\exp\left[-\frac{i}{\hbar}q\bs r \cdot\bs A^\Omega\right]$. It results with:
\begin{equation}
\label{eq:hamsharp}
T^{\Omega} H T^{{\Omega} \dagger}=\frac{\bs p^2}{2m}+V_{\t{Coul}}(\bs r)
\\+q\bs rE^{\Omega}(0)
+E_{\t{dip}}^\Omega+R^\Omega,
\end{equation}
with $R^\Omega=-\frac{q}{m}\bs p\cdot (\bs A(\bs r)-\bs A^\Omega)+\frac{q}{2m}(\bs A(\bs r)-\bs A^\Omega)^2$.
The impact of $R^\Omega$ is assumed to be negligible for the discussed problem (i.e. we assume, that for $|\bk|^2\leq \Omega^2/c^2$ the approximation $e^{i\bs k\bs r}\approx 1$ is accurate on the size of an atom, while for $|\bk|^2> \Omega^2/c^2$ interacting with an atom is negligible, as $\braket{g|e^{i\bs k \bs r}\bs e_{\bk,s}\cdot \bs p |e}$ is small). To calculate $E^{\Omega}_{\t{dip}}$, we go again to continuous space and using Eq. \eqref{eq:cont} obtain: 
\begin{equation}
\label{eq:dipen}
E_{\t{dip}}^\Omega=\frac{q^2\Omega^3}{18\pi^2\epsilon_0c^3}\bs r^2.
\end{equation}
Note that for too large $\Omega$, not only \eqref{eq:dipen} become large compared to the energy gap in the system, but also assumptions about negligibility of $R^\Omega$ fail.

\section{Smooth cutoff}
\label{app:smooth}

The smooth cutoff introduced in section \ref{sec:cutoff} would correspond to the following construction. We introduce the function $f(\bs k,s)$ satisfying $\braket{g|e^{i\bs k \bs r}\bs e_{\bk,s}\cdot \bs p|e}\approx f(\bs k,s)\braket{g|\bs e_{\bk,s}\cdot \bs p|e}$. Therefore $f(\bs k,s)\approx 1$ for large wavelengths and decreases to $0$ for wavelengths much smaller than the size of an atom (in general, it should not be anisotropy). Then we define:
\begin{equation}
\bs A^\epsilon=\sum_{\bk,s}f(\bs k,s)\sqrt{\frac{\hbar }{2\omega_{\bk}V\epsilon_0}}\bs{e}_{\bk,s}\left(-i \hat a_{\bk,s}+h.c.\right)
\end{equation}
and we repeat the procedure from the previous section with 
$T^\epsilon=\exp\left[-\frac{i}{\hbar}q\bs r \cdot\bs A^\epsilon\right]$. We have then 
\begin{equation}
\bs E^\epsilon=\sum_{\bk,s}f(\bs k,s)\sqrt{\frac{\omega_{\bk}\hbar }{2V\epsilon_0}}\bs{e}_{\bk,s}\left( \hat a_{\bk,s}+h.c.\right).
\end{equation}
Next, we assume $f(\bs k,s)$ satisfies 
\begin{equation}
\sum_s\int_0^{2\pi}d\varphi \int_0^\pi d\theta\sin(\theta)\cos^2(\theta) |f(\bs k,s)|^2=\frac{4\pi}{3}e^{-\epsilon c k}
\end{equation}
(the factor $\frac{4\pi}{3}$ comes from the fact that $f(\bs k,s)\overset{|\bs k|\to 0}{\to}1$). It gives:
\begin{equation}
E_{\t{dip}}^\epsilon=\frac{q^2}{3\pi^2\epsilon_0\epsilon^3}\bs r^2.
\end{equation}
As for $1s$ and $2p$ states of hydrogen atom we have $\braket{g|\bs r^2|g}=3a_0^2$, $\braket{e|\bs r^2|e}=30a_0^2$, $\braket{g|\bs r^2|e}=0$, it correspond to the correction to frequency, which for 
$\epsilon\approx 10 a_0/c$, is relatively small when compared to $\nu$. Therefore, after repeating the reasoning from \ref{sec:standard}, one can truly obtain \eqref{eq:main}.

\section{The \texorpdfstring{$\bs A\cdot \bs p$}{TEXT} %
     vs. \texorpdfstring{$\bs E\cdot \bs r$}%
     {TEXT} Hamiltonians}
\label{app:aper}
Looking at all this, relatively complicated, procedure, one may ask a simple question: wouldn't be much more efficient to use $e^{i\bs k\bs r}\approx 1$ for small frequency, but stay in the original representation \eqref{eq:fullham}, in which the term connecting particles coordinates with the modes of the light is $-\frac{q}{m}\bs p \cdot \bs A(\bs r)$? To see the potential problems, consider interaction with a single mode $|\bs k|\ll 1/a_0$ in both representations. Let us name the electromagnetic field operators connected with single plane-waves by:
\begin{equation}
\begin{split}
\bs{A}_{\bs k,s}(\bs r)&=\sqrt{\frac{\hbar }{2\omega_{\bk}V\epsilon_0}}\bs{e}_{\bk,s} \left(-i\hat a_{\bk,s}e^{i\bs k \bs r}+h.c.\right),\\
\bs{E}_{\bs k,s}(\bs r)&=\sqrt{\frac{\hbar\omega_{\bk} }{2V\epsilon_0}}\bs{e}_{\bk,s} \left(\hat a_{\bk,s}e^{i\bs k \bs r}+h.c.\right).
\end{split}
\end{equation}
Using the fact that for Hamiltonian of the form $H_A=\frac{\bs{ p}^2}{2m}+V_{\t{Coul}}(\bs r)$, we have $[H_A,\bs r]=\frac{\hbar}{im}\bs{p}$, so:
\begin{equation}
\braket{g|\bs p|e}=\braket{g|\frac{im}{\hbar}[H_A,\bs r]|e}=-im\nu\braket{g|\bs r|e}
\end{equation}
which leads to:
\begin{equation}
\braket{g;\bk,s|\tfrac{q}{m}\bs p \cdot\bs A_{\bk,s}(0) |e;0}=\frac{\nu}{\omega_k}\braket{g;\bk,s|q\bs r\cdot\bs E_{\bk,s}(0)|e;0}.
\end{equation}
While it gives consistent results close to resonance $\nu\approx \omega_k$, it differs significantly far from it. Note that this effect is not related to breaking conditions $\bs A(\bs r)\approx \bs A(0)$, as going $\omega\to 0$, it is strictly satisfied.
The reason for the difference is that states $\ket{g;\bs k,s},\ket{e;0}$ are not eigenstates of the generator of transformation $T$,
\begin{equation}
\t{span}\{\ket{g;\bk,s},\ket{e;0}\}\neq \t{span}\{T\ket{g;\bk,s},T\ket{e;0}\},
\end{equation} 
therefore the analyses in two representations with the same two-dimensional subspace are not equivalent \cite{rzazewski2004equivalence,cohen1997photons}. That implies the question: for which approach are the states $\ket{g;\bk,s},\ket{e;0}$ more accurate (where $\ket{g},\ket{e}$ are eigenstates of the Hamiltonian $H_A$)?

To clearly distinguish between two approaches, keeping compact notation, from now the original Hamiltonian, in which the interaction term involves vector potential \eqref{eq:fullham}, will be denoted as:
\begin{multline}
H_{\bs{Ap}}:=H=\\
=\frac{1}{2m}(\bs p- q \bs A(\bs r))^2+V_{\t{Coul}}(\bs r)+\sum_{\bk,s} \hbar \omega_{\bk} \hat a_{\bk,s}^\dagger \hat a_{\bk,s},
\end{multline} 
while the one in which the interaction term connects the electron's position with the electric field \eqref{eq:hamsharp} will be named as:
\begin{multline}
H_{\bs{Er}}:=T^{\Omega} H T^{{\Omega} \dagger}=\\
=\frac{\bs p^2}{2m}+V_{\t{Coul}}(\bs r)
-q\bs r\bs E^{\Omega}(0)
+E_{\t{dip}}^\Omega+R^\Omega.
\end{multline} 

In general, more reasonable results are obtained when using the form $H_{\bs{Er}}$. To understand this, let us note that the introduction of the vector potential $\bs A(\bs r)$ into the Hamiltonian changes the relationship between the velocity of the electron and the canonical momentum, while the states $\ket{g},\ket{e}$ have been obtained as the solution of $H_A$ (in absence of electromagnetic field, where velocity is proportional to momentum). 

To see it explicitly, let us calculate the matrix elements for $\ket{g;\bk,s},\ket{e;0}$ in both representations. In $H_{\bs{Ap}}$ we have:
\begin{equation}
\begin{split}
\braket{g;\bk,s|H_{\bs{Ap}}|g;\bk,s}&=\braket{g|H_A|g}+\hbar\omega_k+\frac{q^2\hbar}{2m\omega_k \epsilon_0},\\
\braket{e,0|H_{\bs{Ap}}|e,0}&=\braket{e|H_A|e},\\
\braket{g;\bk,s|H_{\bs{Ap}}|e,0}&=\frac{\nu}{\sqrt{\omega_k}}q\sqrt{\frac{\hbar}{2V\epsilon_0}}\braket{g|\bs e_{\bk,s}\cdot \bs r|e}.
\end{split}
\end{equation} 
For $\omega_k\to 0$ the expectation energy of $\ket{g;\bs k,s}$ becomes even much bigger than for $\ket{e;0}$! Indeed, even the occurrence of a single photon with infinitesimally small energy dramatically changes the relation between canonical momentum and velocity. On the other hand, for $H_{\bs{Er}}$ we have (up to term $R^\Omega$):
\begin{equation}
\begin{split}
\braket{g;\bk,s|H_{\bs{Er}}|g;\bk,s}&=\braket{g|H_A|g}+\hbar\omega_k+\braket{g|E^{\Omega}_{\t{dip}}|g},\\
\braket{e,0|H_{\bs{Er}}|e,0}&=\braket{e|H_A|e}+\braket{e|E^{\Omega}_{\t{dip}}|e}\\
\braket{g;\bk,s|H_{\bs{Er}}|e,0}&=\sqrt{\omega_k}q\sqrt{\frac{\hbar}{2V\epsilon_0}}\braket{g|\bs e_{\bk,s}\cdot \bs r|e},
\end{split}
\end{equation} 
where $\braket{e/g|E^{\Omega}_{\t{dip}}|e/g}$ is $\omega$-independent, relatively small (for proper cutoff) correction. One may also note that for $\omega_{\bk}\to 0$ we recover the proper energy of an electron in a constant electric field.

\end{document}